# Medium Resolution Near Infrared (2.15$\mu$m–2.35$\mu$m) Spectroscopy of Late-type Main Sequence stars

B. Ali[1], John S. Carr[1], D. L. DePoy[1], Jay A. Frogel, and K. Sellgren[1]

Department of Astronomy

The Ohio State University

## ABSTRACT

We present an atlas of moderate resolution ($\lambda/\Delta\lambda$ ~1380) 2.15 $\mu$m to 2.39 $\mu$m spectra of 33 luminosity class V stars of spectral type F3 to M6 in the MK classification system. We find that the equivalent widths of several spectral features vary significantly with temperature and typically allow effective temperature determination accurate to approximately $\pm$350 K.

---

[1]Visiting Astronomer, Cerro Tololo Inter–American Observatory, National Optical Astronomy Observatories, which are operated by the Association of Universities for Research in Astronomy, Inc., under cooperative agreement with the National Science Foundation.

## 1. Introduction

We have obtained moderate resolution (R=1380) 2.15 $\mu$m to 2.39 $\mu$m spectra of 33 luminosity class V stars representing a range in spectral class from F3 to M6 in the MK classification system. These data should have many uses; for example, they begin to provide a basis for spectral classification of pre-main sequence (PMS) stars in obscured young clusters. Also, several places in the Galaxy can only be observed in the infrared due to high extinction, an atlas of representative infrared spectra of main sequence stars can help provide a guide for understanding the underlying population of stars in these areas. A template of stellar spectra in the near-infrared can be used for population synthesis analysis of crowded regions in our own Galaxy (*e.g.* the Galactic center) as well as external galaxies.

Previously, Kleinmann and Hall (1986; hereafter KH86) provided spectra of 26 stars ranging from F8–M7 in spectral class and in luminosity from dwarfs to supergiants. Although the resulting spectra were of relatively high signal-to-noise ratio (S/N) and resolution (R$\sim$3000), the number of stars in each luminosity class was low. In particular, only 6 dwarfs were observed, one of which is a halo star. Other recent works of similar nature include Origlia, Moorwood & Oliva (1993) for the H band and Lancon & Rocca-Volmerange (1992) for the H and K bands; the former did not include dwarfs, and the spectra in the later paper are of modest S/N and resolution.

In this paper we expand on earlier works by presenting moderate resolution, high S/N spectra of a large sample of late-type main sequence stars. In section 5., we discuss how the main absorption features vary with spectral type and in section 6. we compare the measured equivalent widths of these features to predictions based on synthetic spectra computed using model atmospheres. Because our sample is limited to dwarf stars, no attempt will be made to study pressure effects (*i.e.* luminosity classification).

## 2. Selection and Observations of the MK spectral standards

The sample of dwarf spectral standards was taken largely from Keenan and McNeil (1989) and Kirkpatrick, Henry & McCarthy (1991). A few additional stars were added from the Bright Star Catalog (Hoffleit 1982). Where possible we have excluded known halo stars or stars with known metallicity peculiarities. Table 1 lists our final sample. Column 1 lists the star's name, column 2 the MK spectral type, column 3 the spectral type reference, column 4 the date of observation, column 5 the K magnitude, column 6 the exposure time and column 7 the estimated S/N ratio in the spectrum. The S/N is simply the inverse of the rms noise present in our defined continuum regions in our normalized spectra (see section 3.). The K magnitudes are an inhomogeneous collection of values obtained either from the literature or from the visual magnitudes using V-K colours of Bessel and Brett (1988). These K magnitudes are expected to be only accurate to $\sim$0.2 mag and are meant as a guide for calculating exposure times for the objects.

All observations were carried out with the Ohio State InfraRed Imager and Spectrograph (OSIRIS; DePoy *et al.* 1993) on the Cerro Tololo Inter-American Observatory (CTIO) 4 m and 1.5 m telescopes. The slits used were $1''.4 \times 110''$ (1.5m) and $0''.5 \times 40''$ (4 m). To correct for telluric absorption features, an A or B spectral type star was observed as close to the object star airmass as possible. With the exception of the Br$\gamma$ line, these stars are not known to have any features in our observational window at the moderate spectral resolution of the observations.

## 3. Data Reduction and Equivalent Widths

In taking the observations, OSIRIS was first used in imaging mode to acquire the star in the slit. Several spectra were then taken to produce the total exposure time listed in Table 1. During successive exposures the star's position in the slit was moved systematically from the bottom to the top of the slit. This was done to estimate the sky levels in the exposures (see below) and to aid in the removal of fringes present in all OSIRIS spectra. These fringes multiplicatively modulate the observed counts by a typical peak to peak amplitude of $\sim$6% and are rotated $\sim$20 degrees with respect to the rows and columns of the detector. Averaging the spectra of a program star taken at various positions on the slit and dividing by spectra of an atmospheric standard star taken at similar positions one can effectively cancel the effects of fringing down to $<$1% level.

The Image Reduction and Analysis Facility (IRAF[2]) software was used for all data reduction. The reduction process consisted of flat fielding using dome flats, and sky subtraction using a sky frame made by median combination of all data frames of the object. The median combining algorithm rejected the stellar flux values since each spectrum was obtained on a different part of the detector. Bad pixels (dead pixels and cosmic ray hits) were corrected by interpolating across them in both rows and columns and replacing the value by the average of the interpolated values. The data frames were then geometrically transformed to correct for the curvature of the slit image induced by the grating. The geometric transformation was derived from night sky emission lines. Individual spectra from the transformed images were extracted along a 10 pixel wide aperture using the APALL package in IRAF. A further sky subtraction was done by using regions on either side of the aperture. Extracted spectra were normalized and averaged to produce the final spectrum. Spectra of object stars were further divided by atmospheric standards observed and reduced in the same manner to remove telluric absorption features. In ratioing the spectra, we interpolated across the Br$\gamma$ present in the spectra of the atmospheric standards.

[2]The IRAF software is distributed by the National Optical Astronomy Observatories under contract with the National Science Foundation

The presence of the OH air-glow lines (Oliva and Origlia 1992) and the location of the Br$\gamma$ line and the $^{12}$CO(2,0) band heads in our spectra were used to obtain a wavelength solution for the final spectra. Various photospheric absorption features in our spectra were identified by comparing their locations with those of the lines in KH86. The final spectra along with the identified features are shown in Figure 1.

The strongest features evident in our data are listed in Table 2. The equivalent widths of the Na I, Ca I, Mg I features (Table 2; section 5.) and the Brackett $\gamma$ line were measured by defining a global continuum as the best fit line to 6 wavelength regions determined to be free of spectral lines based on the higher resolution spectra in KH86 and the near-infrared solar spectrum of Livingston & Wallace (1991). The band edges for the continuum and the lines are listed in Table 3. These were chosen so as to span the full width of the features of interest. The continuum for the $^{12}$CO(2,0) first overtone bandhead was obtained by adopting the mean flux level of the sixth continuum band (see above) as the continuum over the bandhead region.

The strength of the $H_2O$ vapor absorption was measured as the ratio of the flux level in two continuum bands listed in Table 2.

We have investigated the possible effects on the measured values of the equivalent widths if instruments with resolution different from OSIRIS (R∼1380) were to be used to carry out similar observations. We find that equivalent widths measured from the higher resolution KH86 (R∼3000) digital atlas in the same manner as ours are consistent with our values within the scatter observed in our $T_{\rm eff}$ – equivalent width relationship (see also section 5.1). The halo star Wolf 359, which is likely to be very metal poor, was not included in this comparison. Furthermore, KH86 data was rebinned to the resolution of OSIRIS and the equivalent widths were re-measured. Comparison with equivalent widths derived from full resolution KH86 data shows that the mean difference is 5 $\pm$ 7%. Again, this value is smaller than the observed scatter about our mean $T_{\rm eff}$ – equivalent width relationship (see section 5.1). Therefore, we conclude that observations with spectral resolution between OSIRIS and the KH86 data will not be significantly affected by the instrumental resolution.

To estimate the *formal* errors for the equivalent widths we assumed for the case of background dominated data and weak absorption lines that $\sigma_{\rm line} \sim \sigma_{\rm cont.}$. The error is then given as:

$$\sqrt{2N_{pixels}} \times {\rm dispersion} \times \sigma_{\rm cont}$$

where $N_{pixels}$ = the number of spectral pixels contained within the defined band edges for the line and $\sigma_{\rm cont}$ is the rms noise of the continuum as calculated from the six continuum regions. Errors listed in Table 3 were calculated using this formula. These errors are really lower limits as they provide no estimate for any systematic errors that may exist in the data. For example, uncertainties greater than the formal errors may be due to metallicity variations in our sample (see sec 6.) or inherent differences among effective temperatures of stars of similar spectral types (see sec 4.).

## 4. MK spectral type to effective temperature relationship

The stars in our sample are MK spectral type standards based on empirical criterion assigned for each spectral type. For the purposes of this paper and the interests of the authors, a more meaningful quantity, however, is the star's effective temperature ($T_{\rm eff}$), since $T_{\rm eff}$ measures a physical characteristic of the star. To convert the assigned MK spectral types to effective temperature, we used empirically determined values of effective temperatures for MK standards from 5 different references (Popper 1980; Malagnini & Morossi 1990; Gehren 1981; Bessell 1991; Bell & Gustafsson 1989) as shown in Figure 2. The solid line in the figure is our adopted MK spectral type to effective temperature relationship. It was obtained by fitting a $4^{\rm th}$ order polynomial to the data. For the remainder of the paper, we will adopt the effective temperature as the observed quantity based on this derived relationship.

## 5. Results

The results of our measurements in the form of equivalent width *vs.* spectral type (top axis) and effective temperature (bottom axis) are shown in Figures 3–6. Identification of all of the dominant transitions producing the features we measured are given in Table 2; these were taken from KH86.

### 5.1. The Ca I, Na I and Mg I features

Figure 3 shows the temperature dependence of the Na I doublet, the Ca I triplet and the Mg I features. Values measured for several additional stars from KH86 are plotted as the open triangles for comparision. All of these features arise from the neutral state of the atom and are fine-structure transitions from states having energies $\sim$3–7 eV above ground. Both the Ca I and Na I features have a strong temperature sensitivity as was shown by KH86. Although their sample of dwarfs is much smaller, equivalent widths measured in the same manner show that our data is in agreement with KH86 (see section 3).

Based on modeling results (section 6) and the high resolution near infrared solar atlas of Livingston & Wallace (1991), we find that in stars with $T_{eff}\sim$5800 K the Na I feature is composed of the Na I doublet and an equally strong Si I line (6.73 eV), but the higher excitation Si I line weakens rapidly with decreasing temperature and the Na I lines dominate in the stars cooler than 4500 K. Since the Na I lines have the lowest excitation energy (3.19 eV) of major atomic lines in the three spectral features in Figure 3, the Na I feature is strongest in the coolest stars. Indeed, it is the strongest atomic feature in the K band for stars with $T_{eff}<$3400 K. The Ca I and Mg I features, however, are strongest in the $T_{eff}\sim$3400 K to 5000 K regime because of higher excitation energies (4.68 eV for the Ca I triplet and 6.72 eV for the Mg I line). The Ca I feature is dominated by the Ca I triplet and the 2.263 $\mu$m Fe I line with some contribution from Si I in the hotter ($T_{eff}\sim$5500–6500 K) stars. At our spectral resolution, the Mg I line is blended with a Ca I line and weaker Fe I lines. The Mg I line itself begins decreasing in strength at $T_{eff}\sim$4600 K, but the additional lines in the blend maintain the observed equivalent width of the feature until about

$T_{eff}$~3500 K.

### 5.2. $^{12}$CO(2,0) bandhead

In the spectra of all cool stars the $^{12}$CO(2,0) overtone bandheads are the strongest absorption features in the K band. The measured strength of the bandhead flattens from $T_{eff}$~3500 K to 5000 K (Figure 4). This may be due in part to the increasing importance of $H_2O$ opacity. Our data is consistent with KH86, although here we have defined CO differently, so a direct comparison is difficult. The sensitivity of CO to gravity makes it less useful as a temperature diagnostic, but, as KH86 have shown, the combination of CO and the atomic features can be used as a two dimensional classification.

### 5.3. $H_2O$ absorption

In stars with $T_{eff}$~2700–3500 K, the presence of the $H_2O$ absorption feature can be used to further classify these stars (*e.g.* KH86; Baldwin, Frogel and Persson 1973). In the wavelength region of our observations, the $H_2O$ opacity is strongest near 2.15 $\mu$m region, and weakest around 2.22 $\mu$m region for the coolest stars (Ferriso, Ludwig and Thomson 1966). Thus, the ratio of flux levels in bands centered around these regions can provide a quantitative method for measuring the effect of water on the shape of the dwarf star spectra. Figure 5 shows this ratio as measured in our stars. In stars hotter than $T_{eff}$~3500 K, the presence of Brackett $\gamma$ lines and the weak strength of the $H_2O$ absorption prevents the usefulness of this index. We cannot directly compare our $H_2O$ index with that of KH86 since their second wavelength band lies outside the wavelengths of our observations. Because our $H_2O$ index measures the effect induced on the shape of the spectrum by absorption, it is only useful for stars with low reddening, since reddening also changes the measured slopes of spectra. However, for the typical interstellar reddening law, $A_V \geq 10$ mag is needed to produce a 5% change (from 1.05 to 1.0) in the $H_2O$ index value. This change systematically increases $T_{\rm eff}$ to higher values. We do not expect that any of the stars we observed are reddened by more than $A_V \sim 5$ mag (based on optical colours of the program stars). Therefore, reddening effects do not significantly alter our measured $H_2O$ indices.

### 5.4. H I Br$\gamma$

The H I Brackett $\gamma$ line is the strongest atomic feature in the 2.2 $\mu$m atmospheric window in early type stars. For convenience we show the Brackett$\gamma$ equivalent width measured in all our stars, including those we used for atmospheric calibration, in Figure 6. Note that the relatively small change in the Brackett $\gamma$ equivalent width for B, A, and early F type stars suggests that accurate determination of $T_{\rm eff}$ will be difficult based solely upon Brackett $\gamma$ line strength. Additionally, as shown by Elias (1978), the Brackett $\gamma$ equivalent width peaks near $T_{\rm eff} \sim 9500$ K and becomes double valued for equivalent widths from 4-8 angstroms, thus making it even more problematic for effective temperature determinations.

## 6. Modeling of synthetic spectra

We have calculated equivalent widths from synthetic spectra for comparison with our measured equivalent widths in order to study the behavior of the spectral features with temperature and metallicity. We used a new version of the synthesis program MOOG (Sneden 1973) to predict the blend equivalent width of the Na I, Ca I and Mg I features we see in our data. For model atmospheres, we used a grid of Kurucz models (Kurucz 1992) which span 3500 K to 6750 K in effective temperature at various values of metallicity [m/H] and log $g$. Since the Kurucz models do not include water vapor opacity, results for $T_{\rm eff} \leq 4000$ K are not included. Our initial line lists were taken from the atomic and molecular line lists of Kurucz (1994). This list was used to generate synthetic spectra for the Sun over the wavelength regions of interest, using the Kurucz solar model (Kurucz 1991) with a constant microturbulence $\xi$ = 1.5 km s$^{-1}$ and input solar abundances from Grevesse & Anders (1989). Next, the synthetic spectra were compared to the high-resolution near-infrared solar intensity atlas of Livingston & Wallace (1991). The $gf$ values were adjusted for individual lines to match the observed strengths and spurious or unobservable lines were deleted from the list. For the strongest lines, the damping constants were also adjusted to fit the line profile wings. In order to account for any low-excitation lines not present in the solar spectrum, synthetic spectra were calculated for Arcturus using the model and abundances given in Peterson, Ore & Kurucz (1993), and the results were compared to an archival KPNO 4m FTS spectrum (kindly provided by K. Hinkle). CN lines were excluded from the final list since these have negligible strength in the dwarfs under consideration. Blend equivalent widths were calculated for each of the three spectral features for each temperature of the Kurucz grid with log $g$ = 4.5 and $\xi$ = 2.0 km s$^{-1}$ . The effect of metallicity was investigated by running models for [m/H] = $-0.3$ and $-1.0$.

These results are compared to the observed equivalent widths in Figure 3. In general, the predicted equivalent widths and trends show rough agreement with the data. The observed difference between the measured and calculated values for Na I and Ca I for stars with $T_{\rm eff}$ between 5000 K to 7000 K are consistent with the measured uncertainties and [m/H] variations of $\sim$0.3 dex. The models are not useful below 4000 K, but examination of the predicted span of Na I equivalent width values for stars with $T_{\rm eff}$ in the range 3700 K to 5000 K suggests that a spread in metallicity of $\sim$1 dex could explain the deviations for stars with $T_{\rm eff}$<3400 K. This spread is consistent with the range of [Fe/H] = $-1$ to $+0.2$ observed by Edvardsson et al. (1993) for F and G disk dwarfs.

## 7. Discussion

In order to discuss the use of equivalent width – $T_{\rm eff}$ correlations to classify stars, we adopted mean equivalent width – $T_{\rm eff}$ relationships by fitting low order polynomials to the data. These are illustrated in Figures 7, 8, & 9. Because of the double valued nature of the Ca I feature about $T_{\rm eff} \sim 3500$ K, two fits were needed. However, it should not be difficult to decide on which side of the Ca I curve a particular spectrum lies if Na I and CO equivalent widths are also available. The relationships thus obtained allow one to determine the effective temperature of a star simply by evaluating the polynomial fits with the measured equivalent width of the feature. The fit co-efficients are listed in Table 4.

We estimated the error associated with the resulting value for $T_{\rm eff}$ by *reclassifying* our program stars using these mean relationships and defining the error as the mean deviation found between the adopted effective temperature of the star (from section 4.) and the reclassified temperature. Resulting values for errors are given in Table 5. Typical values for $T_{eff}$ derived from our scheme will be accurate to approximately $\pm 350$ K.

*Formal* errors calculated for the equivalent width measurements themselves (see section 3.; Table 3) do not provide a good way to estimate error in the $T_{eff}$, since the scatter about our adopted mean relationships is dominated by systematic uncertainties such as metallicity variations in our sample (section 6.). The *formal* errors in the equivalent width determination do not account for these systematic effects. For low S/N spectra, however, the formal errors may dominate in the determination of the effective temperature. Thus, in general, one should adopt the larger of the two estimates for the error in $T_{eff}$.

The excitation and ionization potentials for a particular transition and the atom dictate the range in $T_{eff}$ over which a particular feature is useful. For instance, Na I atoms have low ionization potentials and ionize easily in the atmospheres of all but the coolest stars observed. Na I, therefore, provides excellent estimates for stars cooler than $T_{eff} < 6000$ K. The Ca I feature, because of its slightly higher excitation energy, tends to disappear for the very cool stars observed and, therefore, works best for stars with $T_{eff} = 3000$ to 6600 K. Similarly, owing to the dissociation of the CO molecule, the $^{12}CO(2,0)$ bandhead feature becomes undetectable with our resolution and S/N for dwarfs hotter than $T_{eff} \sim 5800$ K. Finally, as stated earlier (see section 5.), the $H_2O$ index is useful only for stars with $T_{eff} < 3500$ K.

We also explored combinations of feature strengths (*e.g.* Na I + Ca I) as effective temperature diagnostics. Results, however, showed that composite indices did not improve the error beyond those listed in Table 4. The scatter in Mg I and Brackett $\gamma$ line was considered too large to be useful for an effective temperature diagnostic.

Since it is likely that most of the scatter over the observational errors in Figures 7 and 8 can be accounted for by abundance variations among the sample stars, the use of such mean equivalent width – temperature relationships to classify stars could introduce errors. For example, if the mean metallicity of our standard star sample is similar to the solar neighbor F-G stars in Edvardsson *et al.* (1993), [Fe/H] $\sim -0.2$, than a systematic offset of $\sim$ 250 K would be introduced in classifying a population of solar metallicity stars. (We note that abundance determinations of T Tauri stars indicate [Fe/H] $\sim 0$ in nearby star-forming regions, Padgett 1992). Even larger errors are possible if a very small sample of stars, such as the KH86 dwarfs, is relied upon to establish the calibration. It should, therefore, be noted that our adopted mean relationships are applicable only to stars that span a similar metallicity range as the stars in our sample.

## 8. Conclusions

Through a sample of 33 dwarf spectral standards, we have shown that spectral features exist in the near infrared K atmospheric window that can be used to estimate the effective temperature (or spectral class) of the stars with moderate resolution spectroscopy. The general behavior seen in our data agrees with the predicted values from synthetic spectra calculated from model atmospheres. However, as evident from our data and modeling, the use of observed equivalent widths of standards without accounting for metallicity can lead to significant errors. More accurate $T_{eff}$ diagnostics can be obtained if the metallicity of the program stars can be independently estimated. Future work will expand upon this study by investigating the effects of metallicity and gravity on the spectral features discussed here.

We would like to thank Bob Wing and Rick Pogge for providing useful comments and discussion on the subject. JAF's research on stellar spectroscopy is supported by NSF grant AST92-18281. OSIRIS was built using funds from NSF grant AST 90-16112 and AST 92-18449. JSC acknowledges support from a Columbus Fellowship at the Ohio State University.

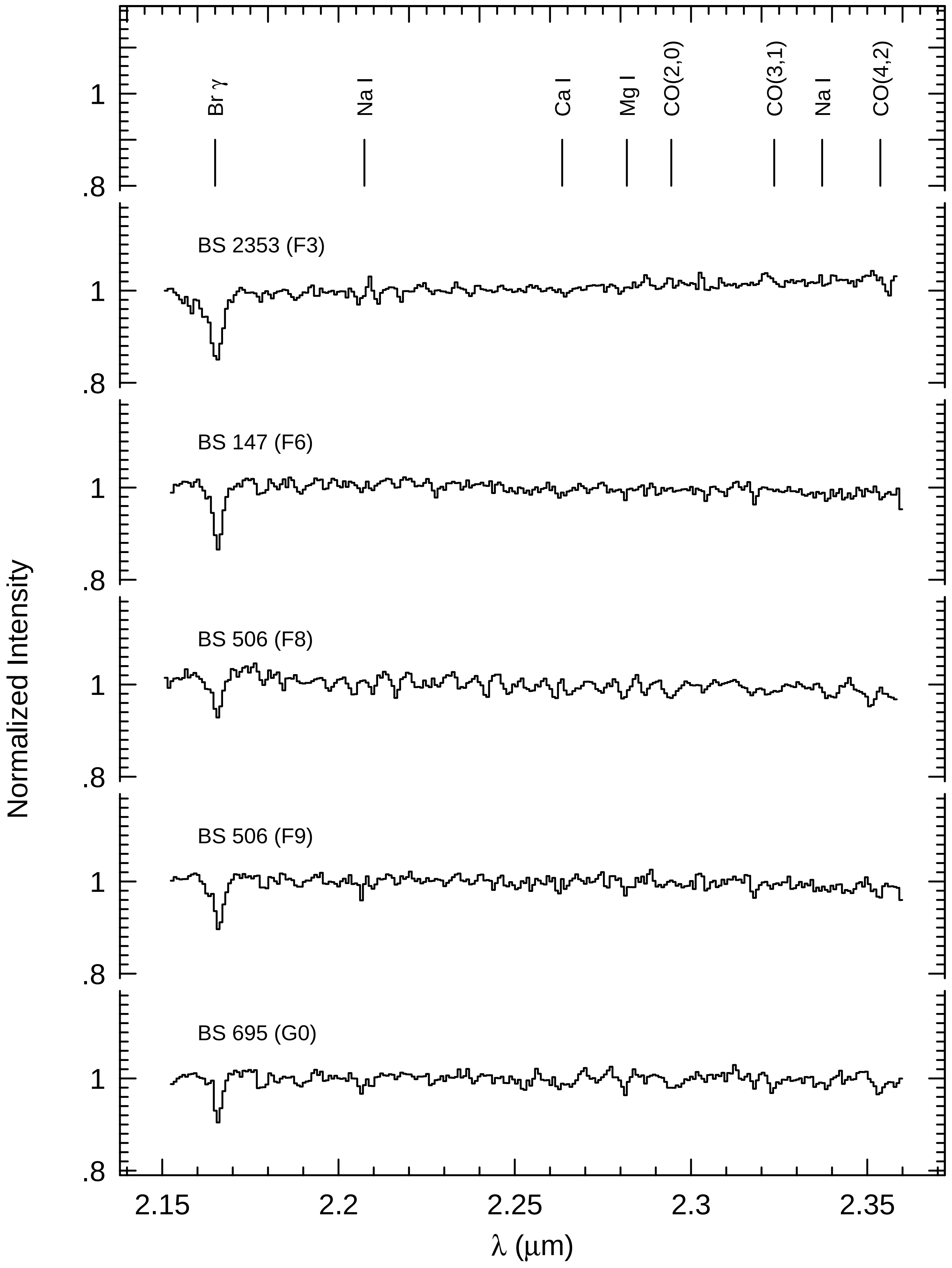

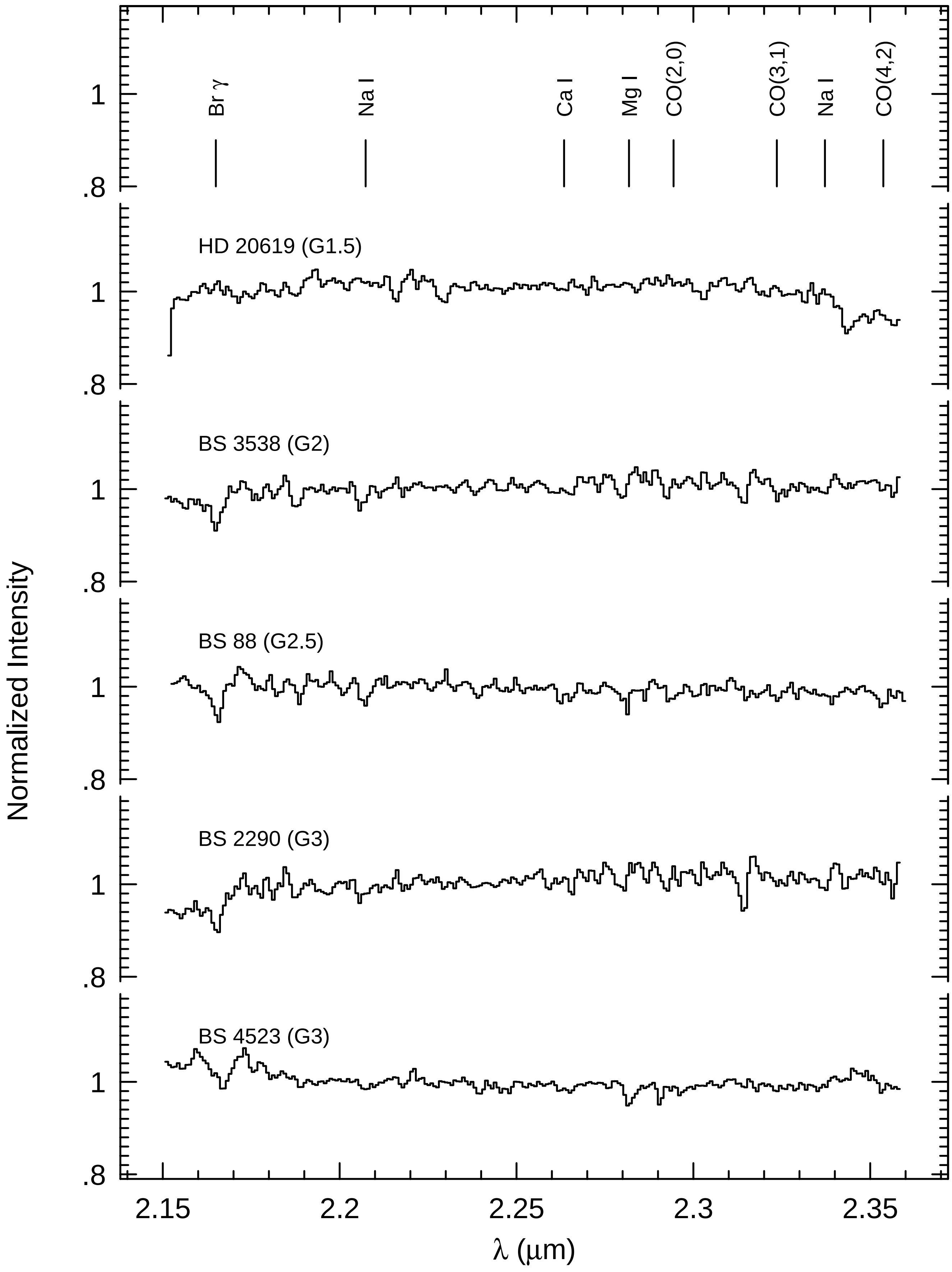

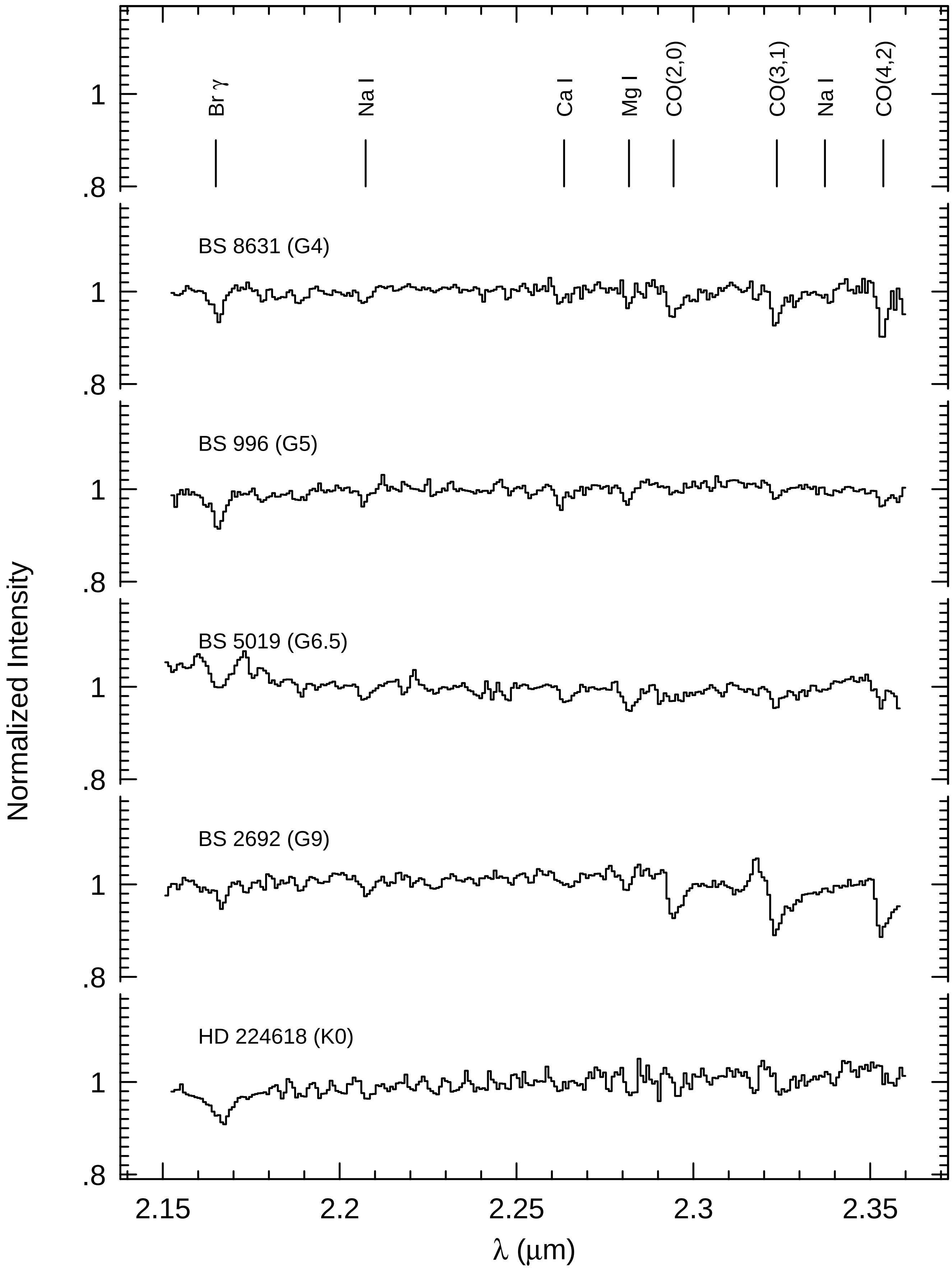

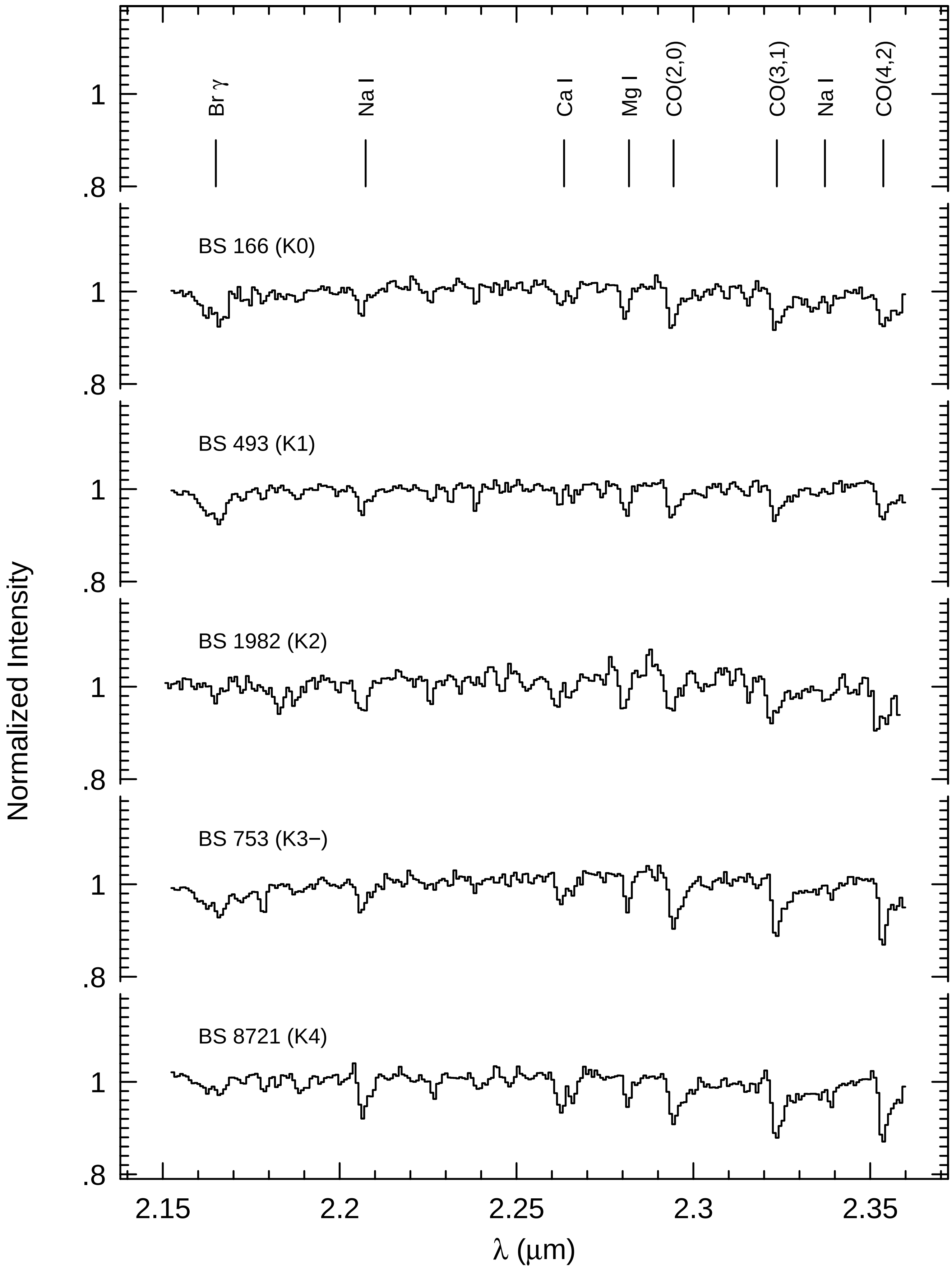

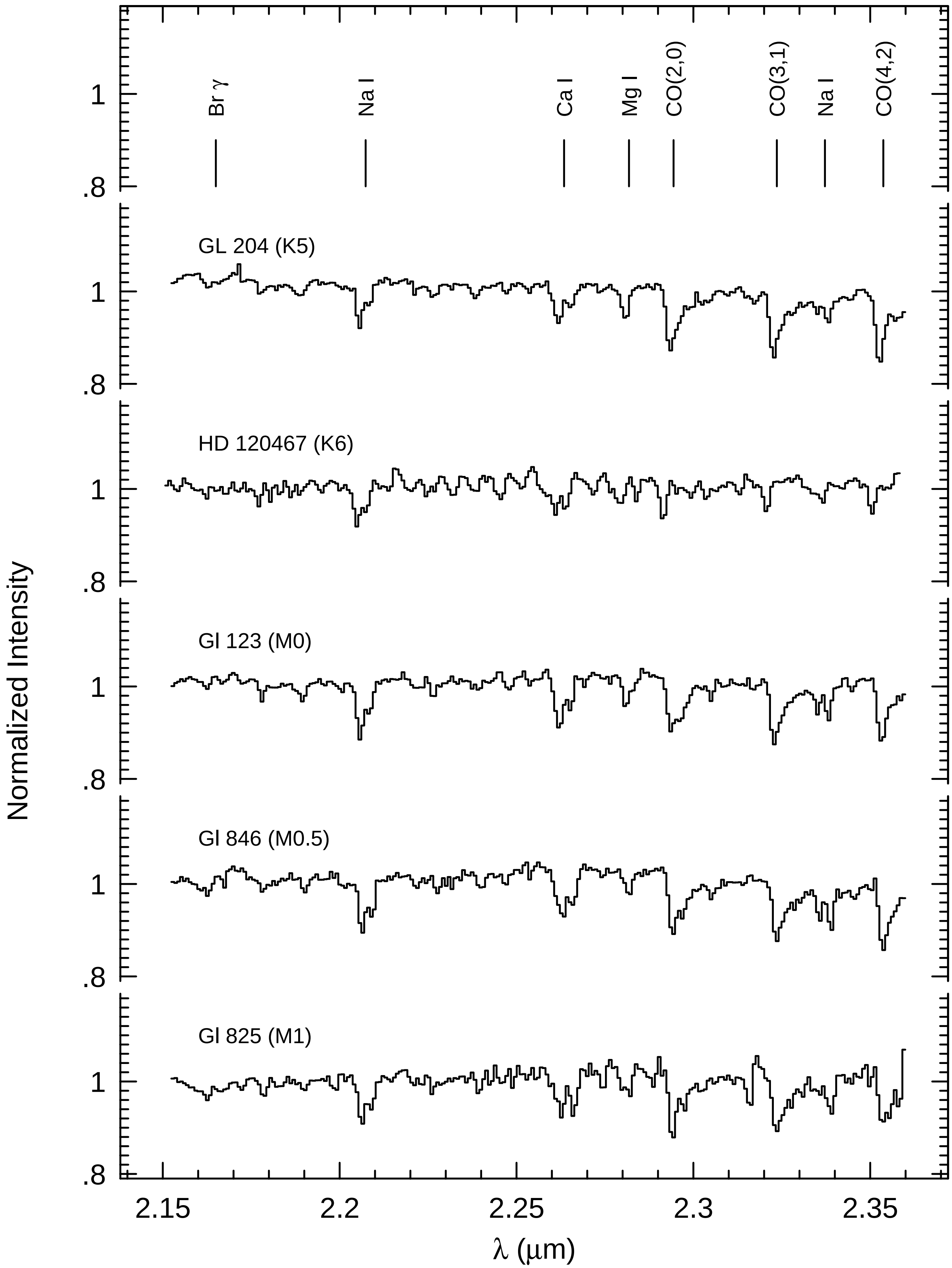

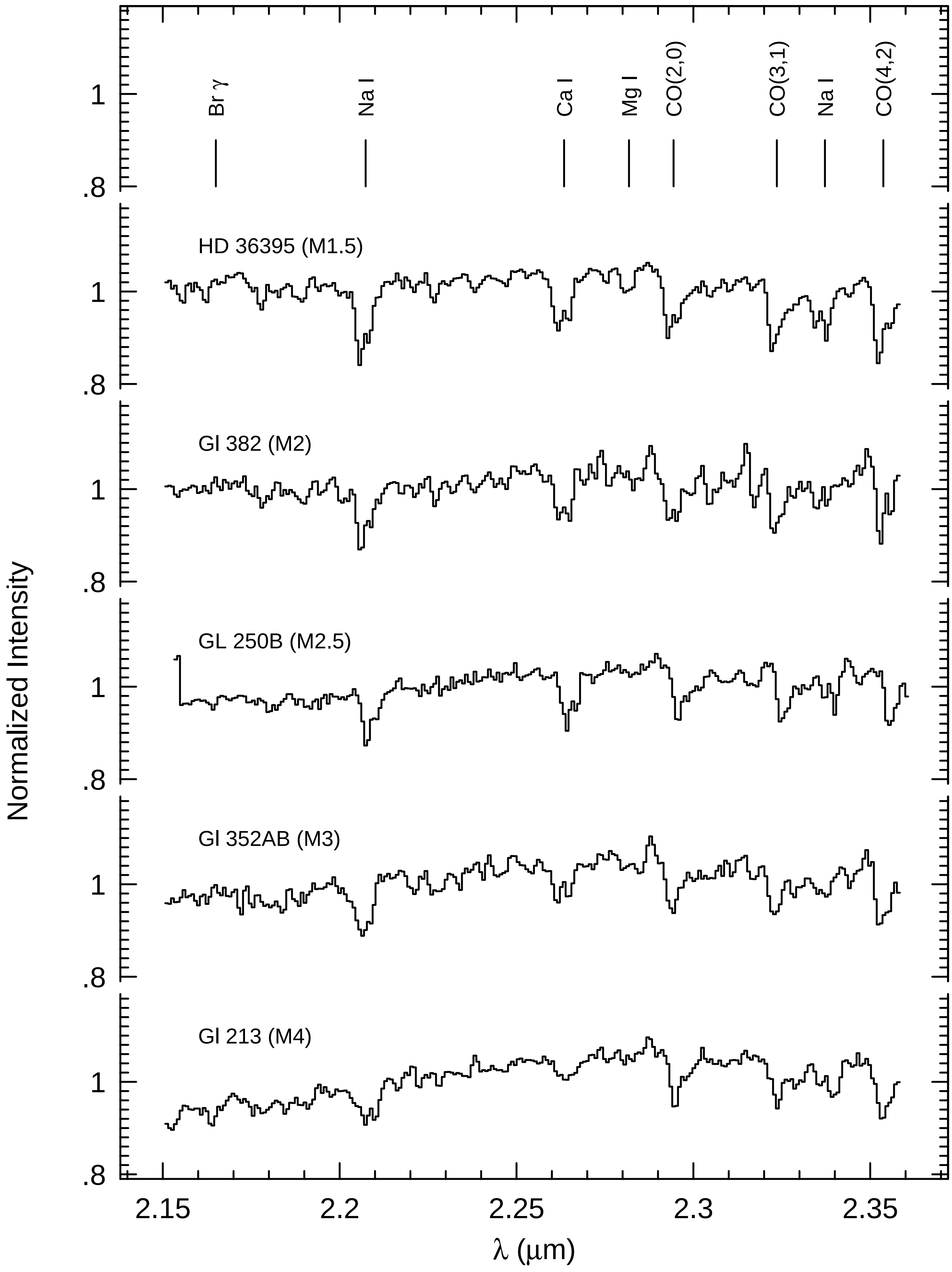

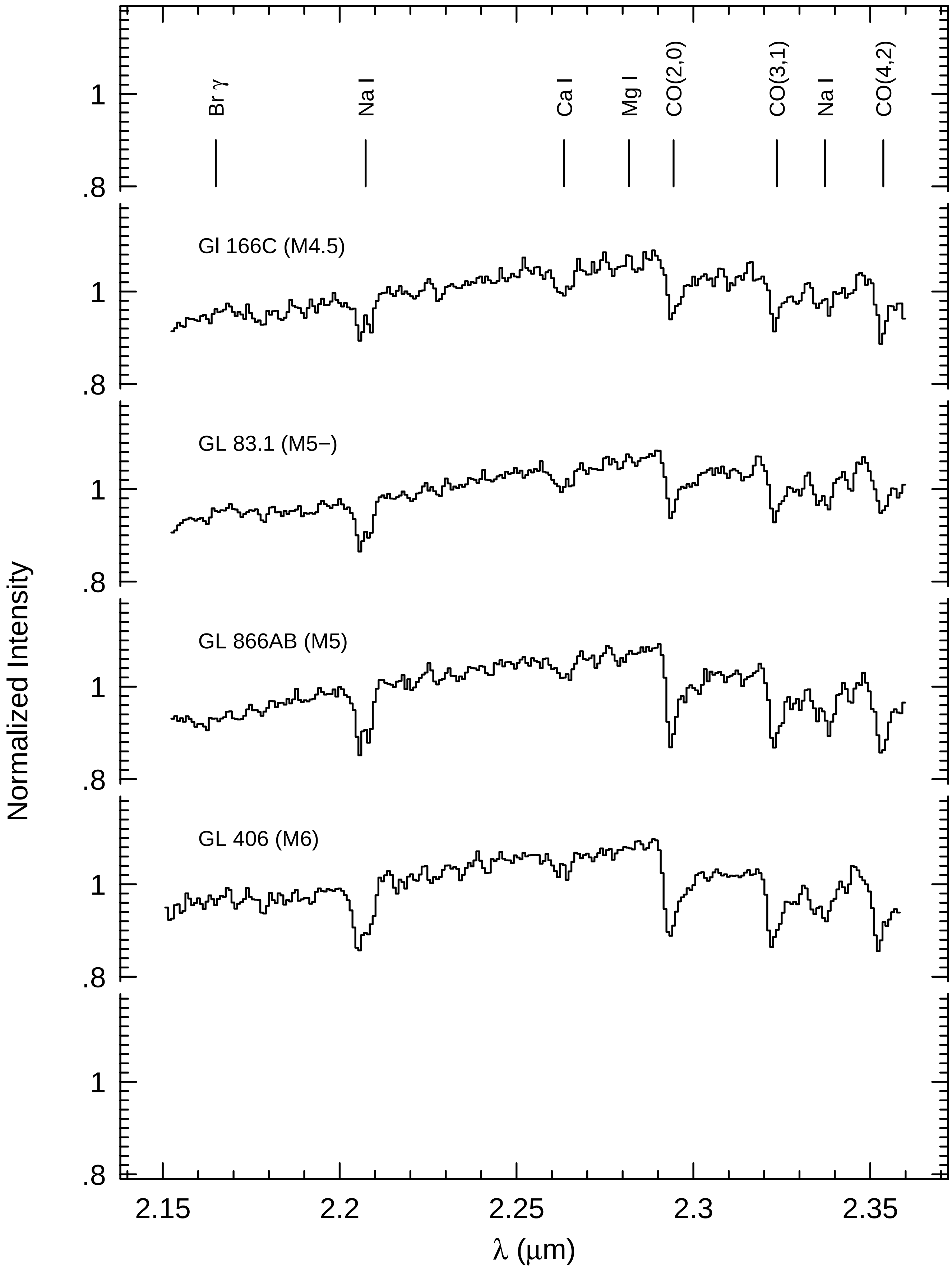

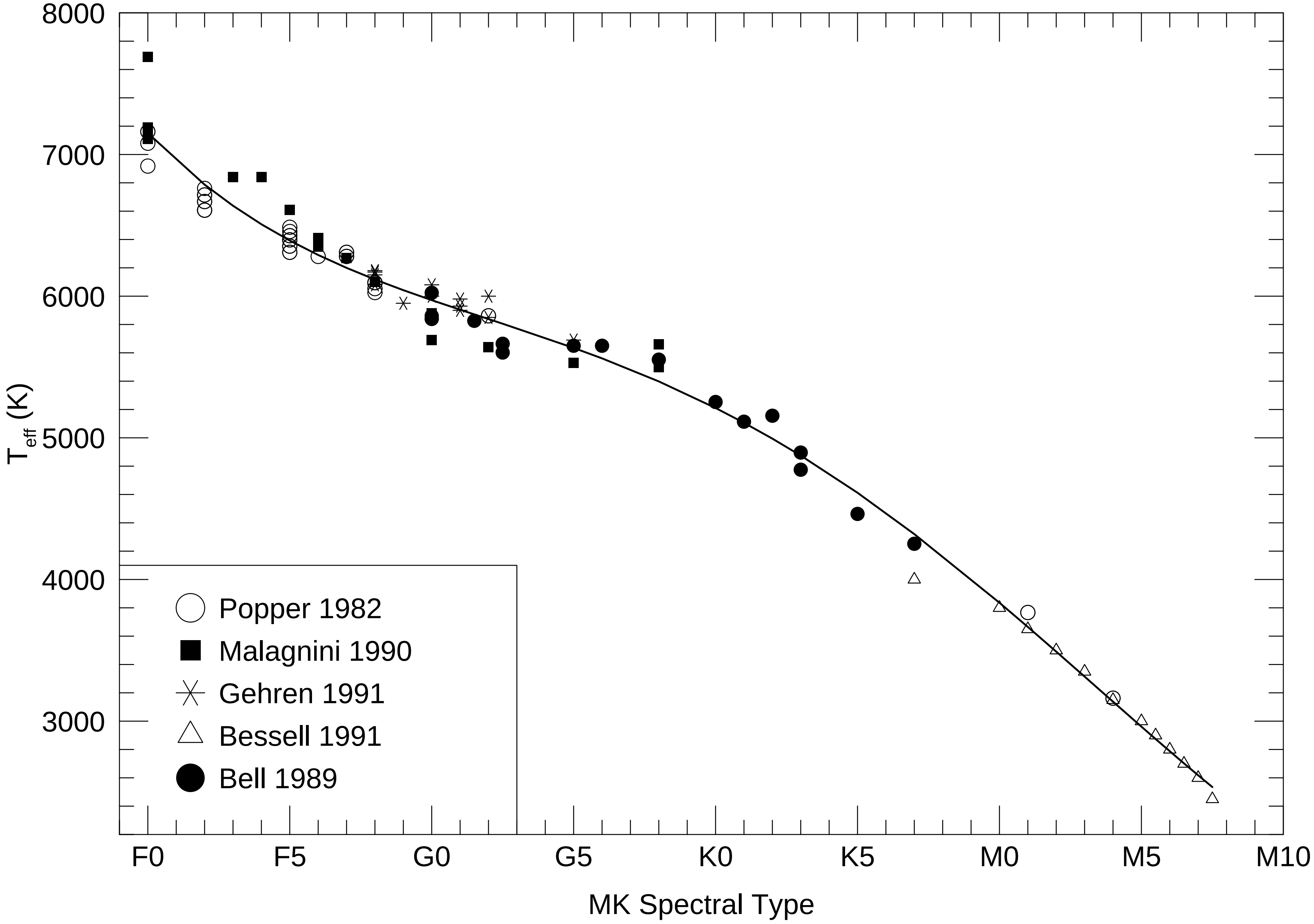

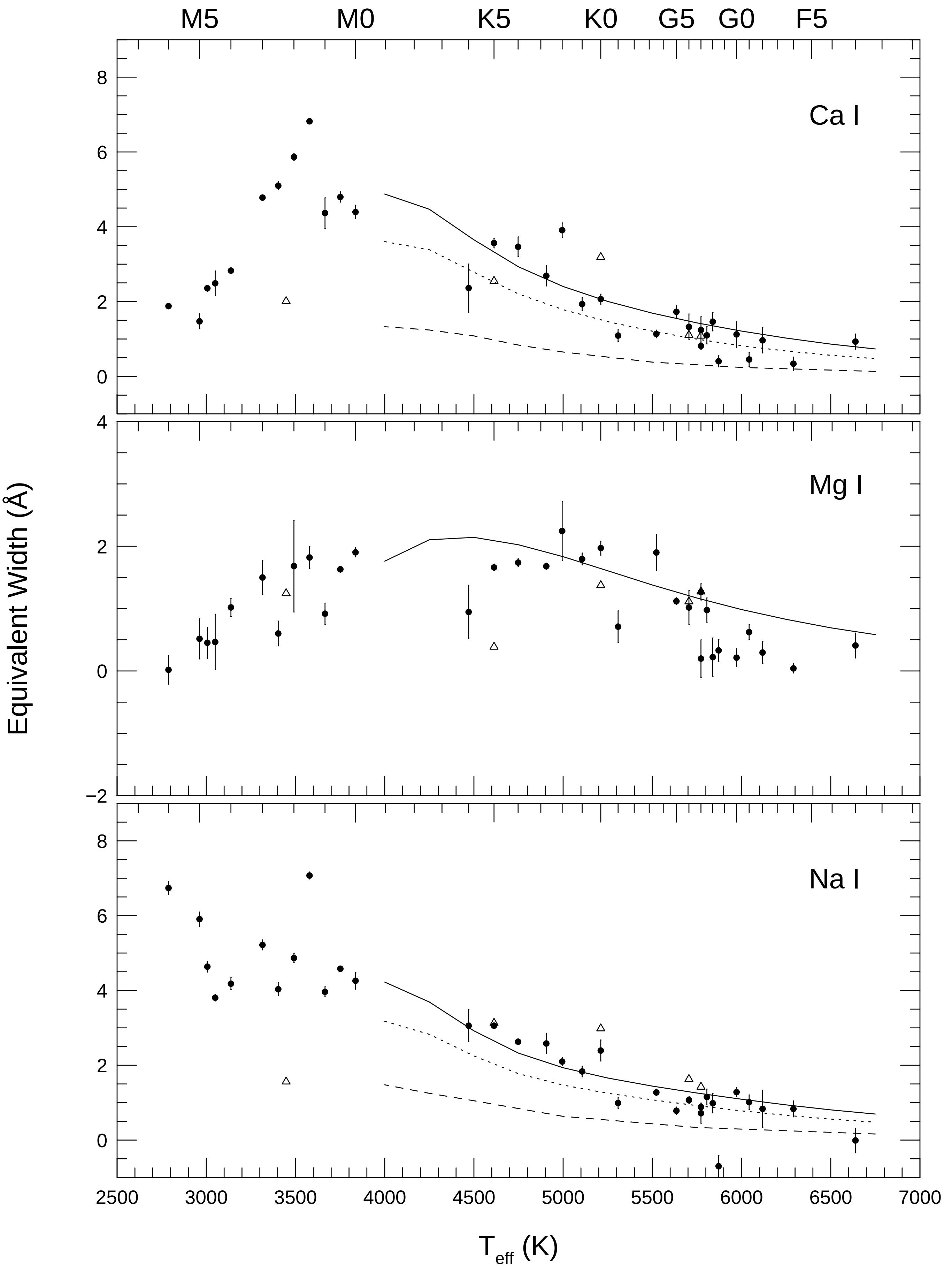

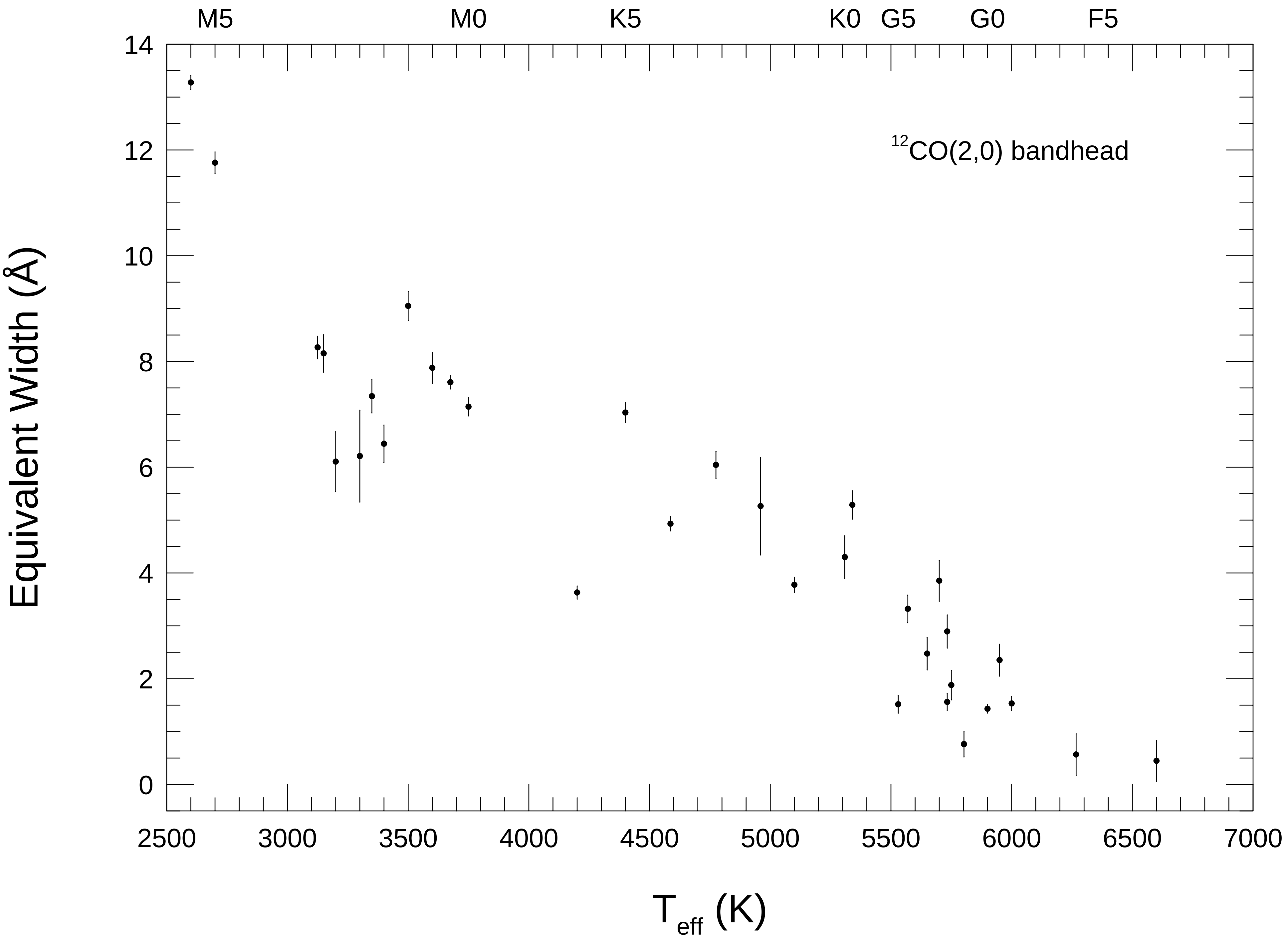

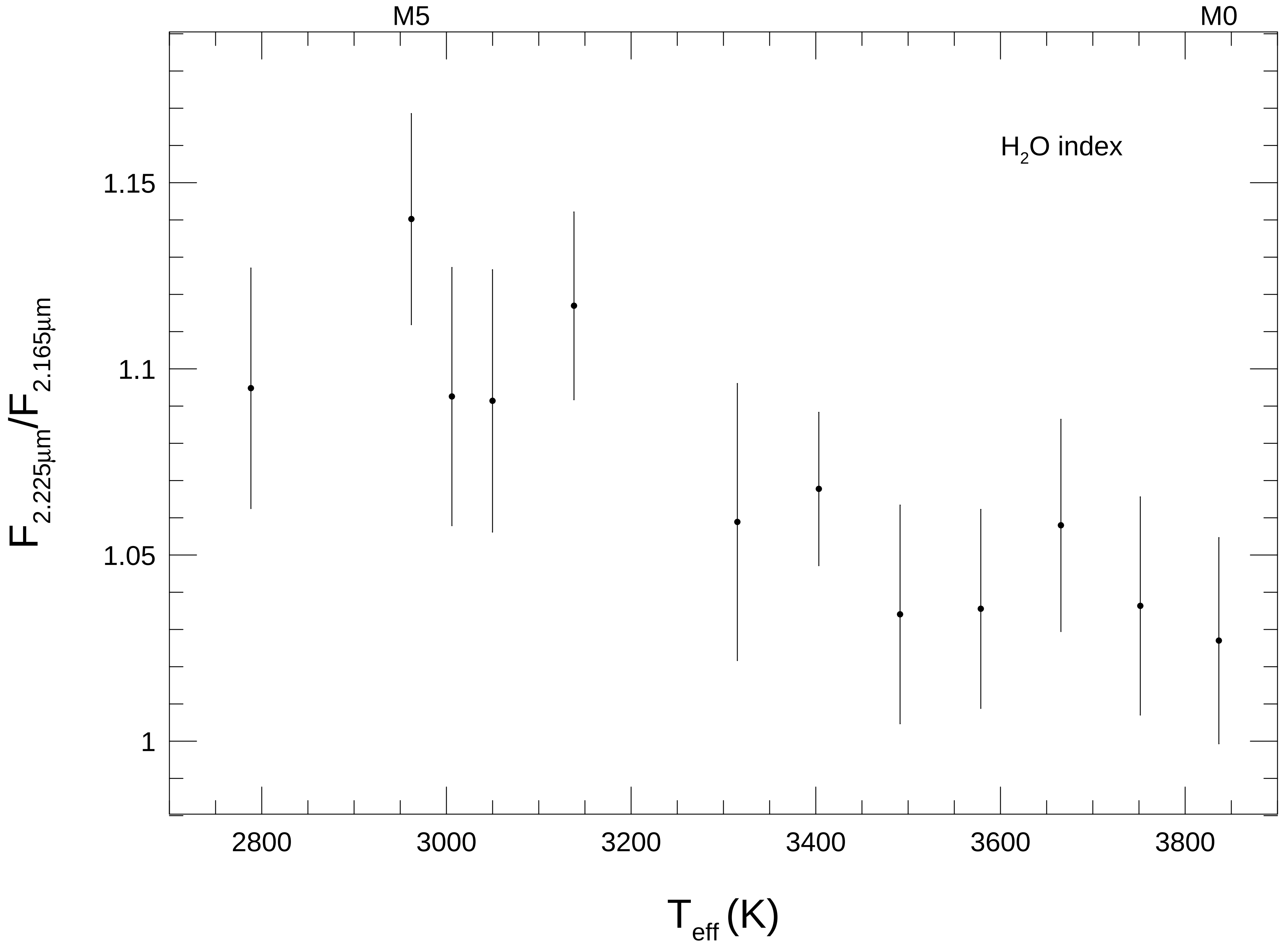

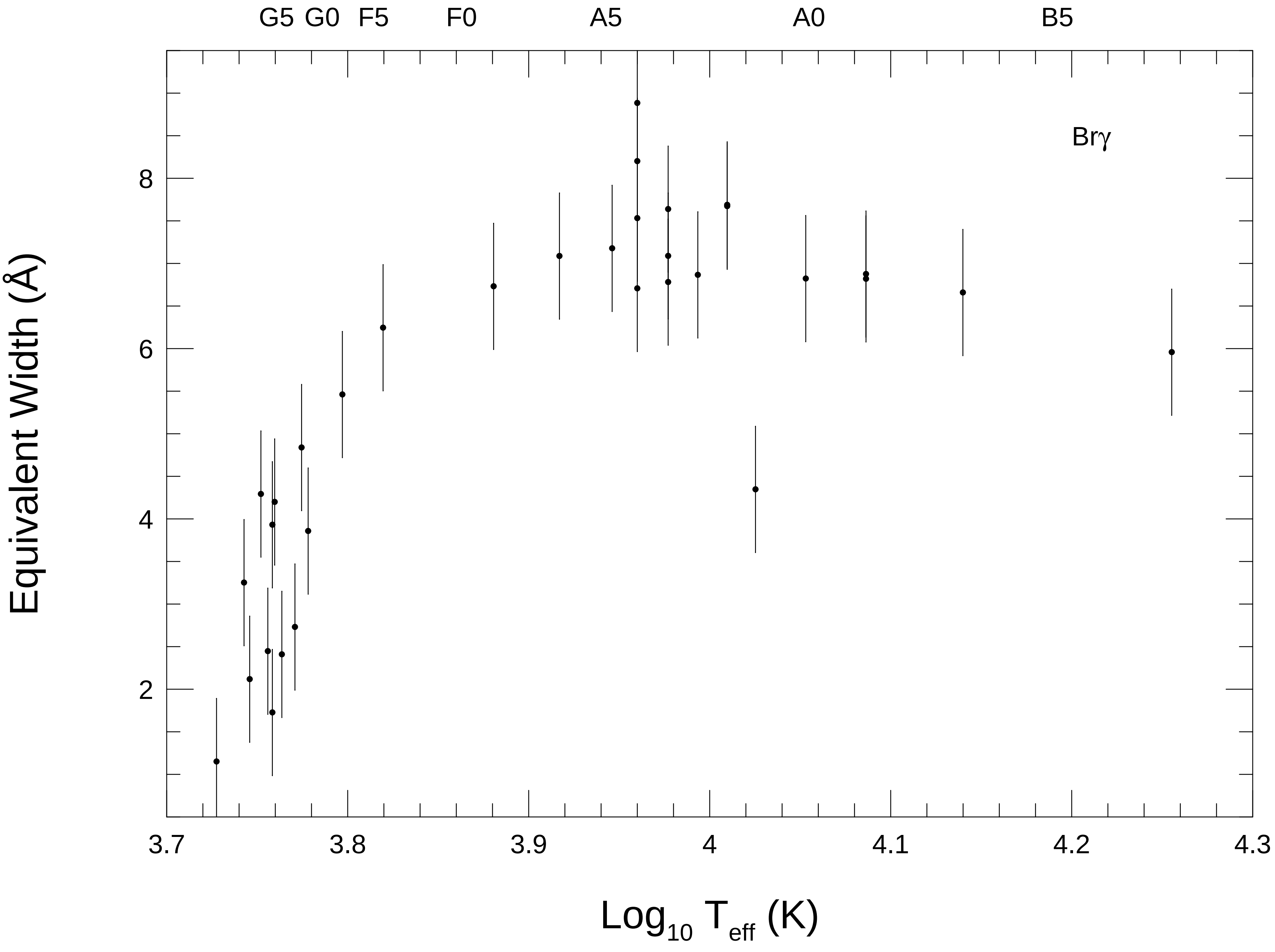

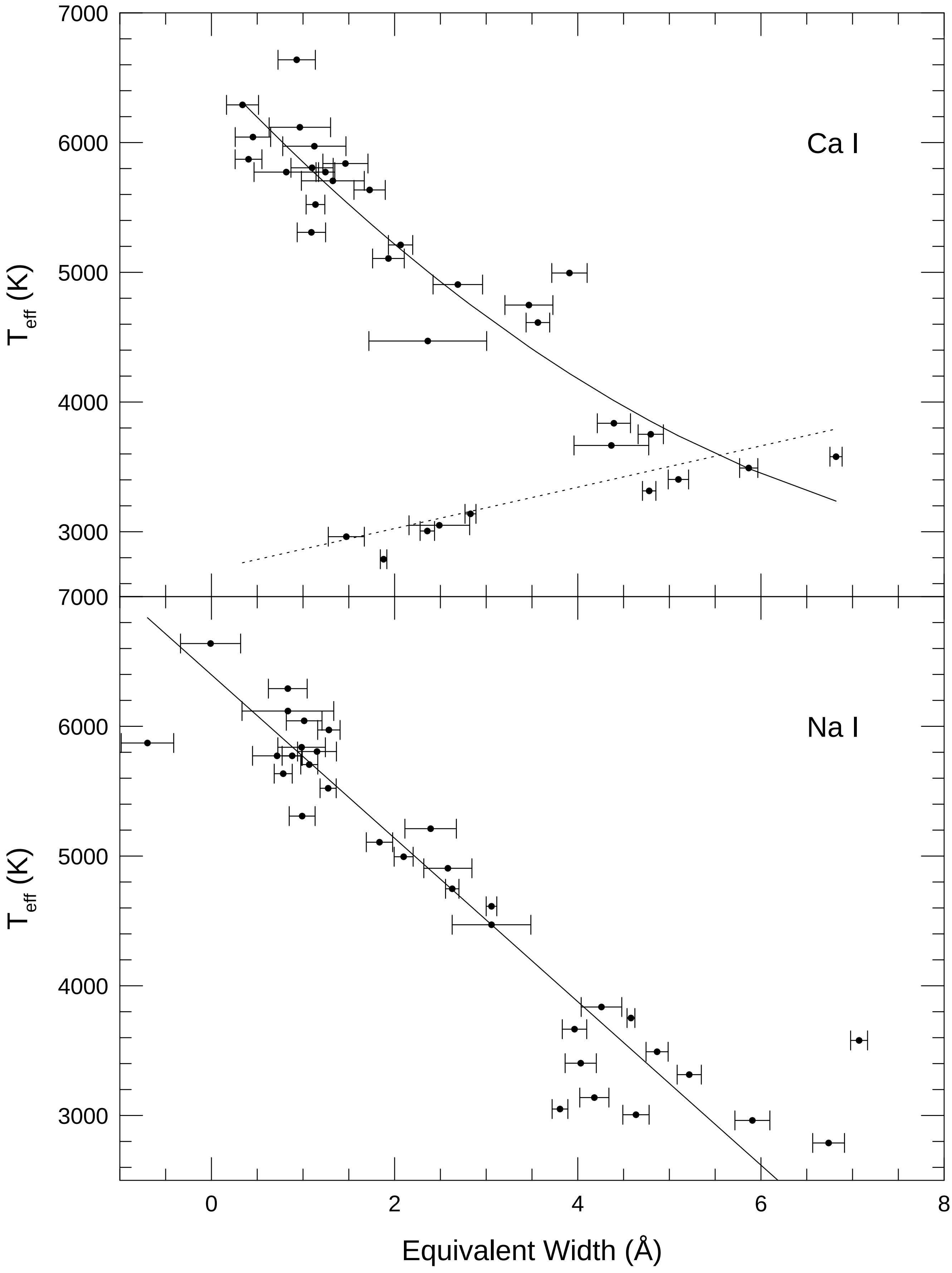

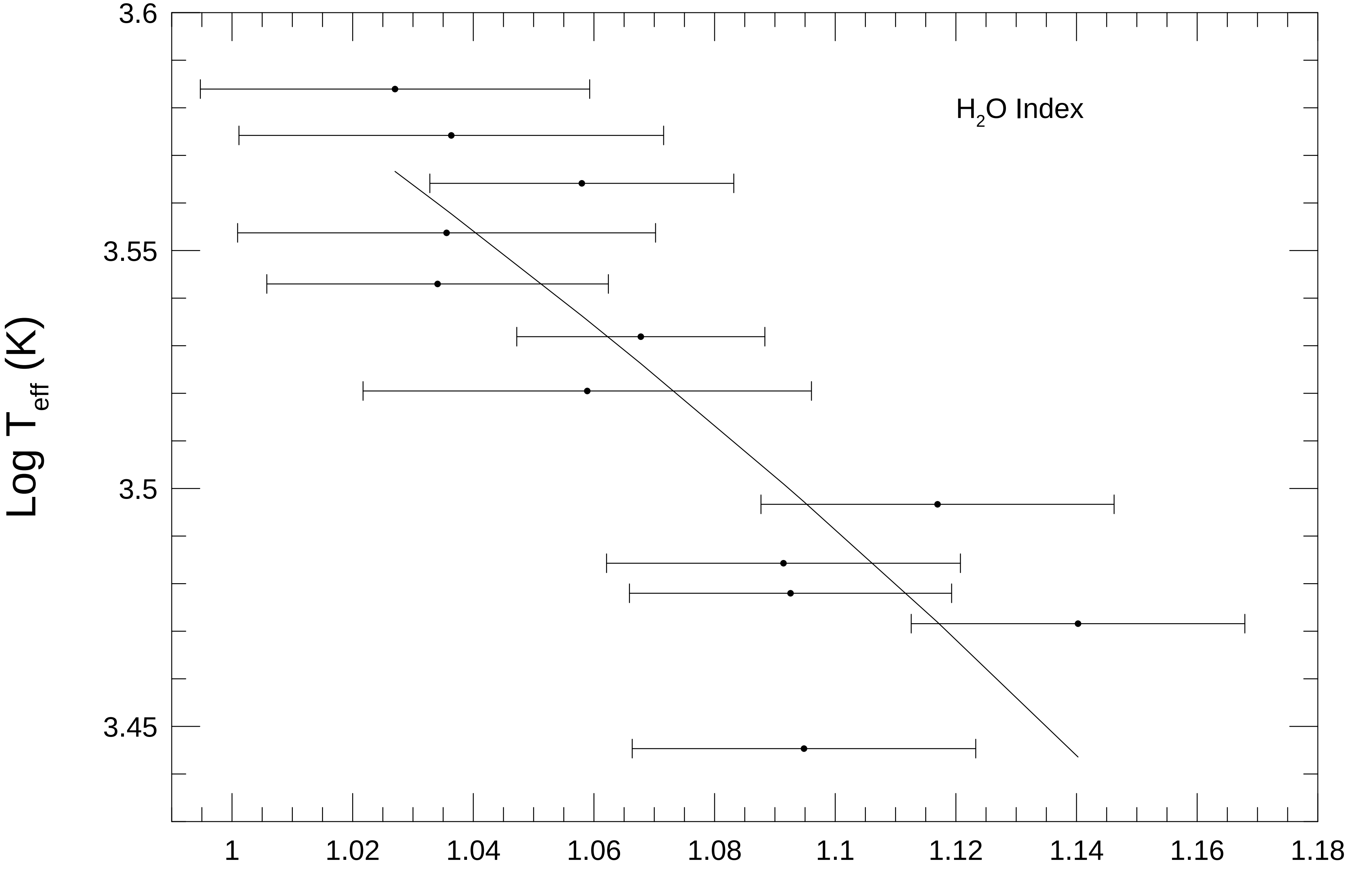

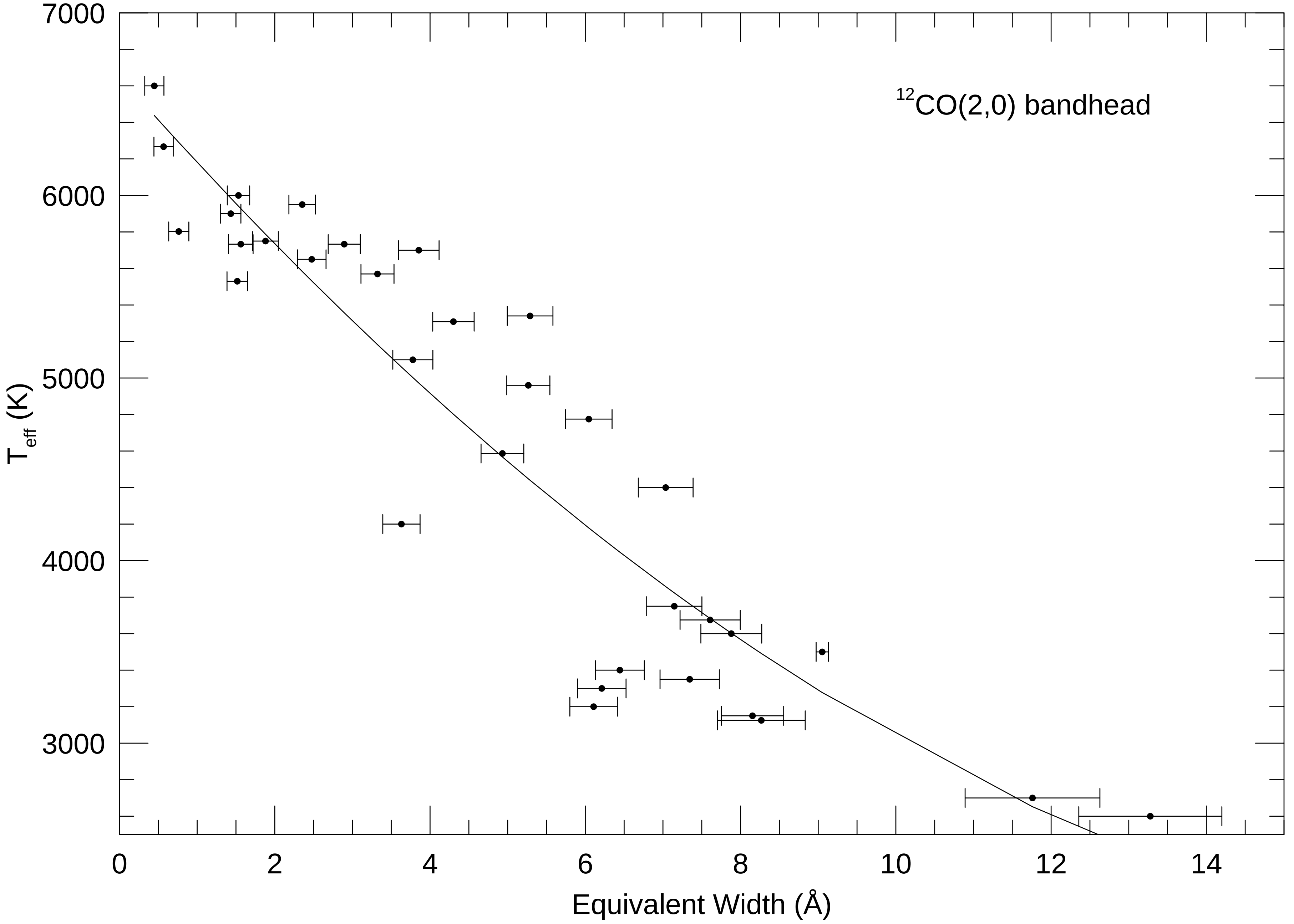

## Figure Captions

**Figure 1:** Normalized spectra of the program stars. Each star is identified in the upper left hand corner of the plot with its spectral class in parentheses. All prominent features are identified at top. For the features discussed in text, the vertical bar at top is shown at the central wavelength of the integration region.

**Figure 2:** Empirically determined effective temperature and spectral types from five different references for main sequence stars. The solid line represents the results of a fourth order polynomial fit to the data.

**Figure 3:** The measured equivalent widths of the spectral type diagnostic features Ca I, Mg I and Na I plotted against the star's effective temperature. The open triangles are equivalent widths measured from the KH86 digital atlas. The lines represents predicted line strengths from synthetic spectra (see section 6). Solid lines are for [Fe/H]=0.0, dotted lines for [Fe/H]=−0.3 and dashed lines for [Fe/H]=−1.0. Definition of the measured band strengths are given in Table 2.

**Figure 4:** The measured equivalent widths of the spectral type diagnostic feature $^{12}$CO(2-0) plotted against the star's effective temperature. Definition of the measured band strength is given in Table 2.

**Figure 5:** The measured strength of the $H_2O$ index (as defined in Table 2) for the late type dwarfs.

**Figure 6:** The measured equivalent widths for the H I Brackett $\gamma$ line plotted against the star's effective temperature. Definition of the measured band strength is given in Table 2.

**Figure 7:** The Na I and Ca I feature and the adopted mean equivalent width to effective temperature relationship (solid and dashed lines) are shown. The average scatter about the adopted fit is $\pm300$ K for the Ca I feature and $\pm450$ K for the Na I feature.

**Figure 8:** The $^{12}$CO(2,0) bandhead equivalent widths and the adopted mean

relationship (solid line) are shown. The average scatter about the adopted fit is ±450 K.

**Figure 9:** The $H_2O$ index and the adopted mean relationship (solid line) are shown. The average scatter about the adopted fit is ±200 K.

Table 1
The observed sample of dwarf stars

| Name | Spectral Type | ref. | Date of Obs. | $K$ mag | $T_{exp.}$ (sec.) | S/N est. |
|---|---|---|---|---|---|---|
| BS 2353 | F3 V | 3 | 25 Jan 94 | 5.6 | 360 | 140 |
| BS 147 | F6 V | 3 | 28 Nov 93 | 4.9 | 15 | 140 |
| BS 2313 | F8 V | 3 | 25 Jan 94 | 5.2 | 195 | 90 |
| BS 506 | F9 V | 3 | 28 Nov 93 | 4.9 | 15 | 100 |
| BS 695 | G0 V | 1 | 28 Nov 93 | 4.0 | 15 | 110 |
| HD 20619 | G1.5 V | 1 | 25 Jan 94 | 5.5 | 210 | 90 |
| BS 2290 | G2 V | 3 | 25 Jan 94 | 5.1 | 360 | 70 |
| BS 3538 | G2 V | 1 | 25 Jan 94 | 4.6 | 195 | 90 |
| BS 88 | G2.5 V | 1 | 27 Nov 93 | 4.9 | 15 | 110 |
| BS 4523 | G3 V | 3 | 26 Jan 94 | 3.4 | 66 | 170 |
| BS 8631 | G4 V | 1 | 28 Nov 93 | 4.2 | 15 | 130 |
| BS 996 | G5 V | 1 | 28 Nov 93 | 3.3 | 17 | 150 |
| BS 5019 | G6.5 V | 1 | 26 Jan 94 | 3.1 | 66 | 170 |
| BS 2692 | G9 V | 3 | 25 Jan 94 | 4.6 | 195 | 130 |
| BS 166 | K0 V | 1 | 27 Nov 93 | 3.9 | 15 | 120 |
| BS 493 | K1 V | 1 | 27 Nov 93 | 3.2 | 15 | 160 |
| BS 1982 | K2 V | 3 | 25 Jan 94 | 3.9 | 26 | 80 |
| BS 753 | K3 V | 1 | 27 Nov 93 | 3.4 | 15 | 130 |
| BS 8721 | K4 V | 3 | 27 Nov 93 | 3.9 | 15 | 150 |
| HD 120467 | K6 V | 1 | 25 Jan 94 | 5.1 | 360 | 60 |
| GL 204 | K5 V | 1 | 28 Nov 93 | 4.6 | 108 | 170 |
| GL 123 | M0 V | 1 | 28 Nov 93 | 5.8 | 84 | 150 |
| GL 846 | M0.5 V | 1 | 27 Nov 93 | 5.8 | 15 | 160 |
| GL 825 | M1 V | 1 | 27 Nov 93 | 3.2 | 14 | 80 |
| HD 36395 | M1.5 V | 1 | 25 Jan 94 | 4.3 | 120 | 110 |
| GL 382 | M2 V | 2 | 25 Jan 94 | 5.1 | 180 | 50 |
| GL 250B | M2.5 V | 2 | 28 Nov 93 | 5.4 | 90 | 90 |
| GL 352AB | M3 V | 2 | 25 Jan 94 | 5.6 | 390 | 70 |
| GL 213 | M4 V | 2 | 25 Jan 94 | 6.4 | 360 | 100 |
| GL 166C | M4.5 V | 2 | 28 Nov 93 | 6.0 | 108 | 90 |
| GL 83.1 | M4.5 V | 2 | 28 Nov 93 | 6.7 | 96 | 120 |
| GL 866AB | M5 V | 2 | 27 Nov 93 | 7.8 | 34 | 100 |
| GL 406 | M6 V | 2 | 25 Jan 94 | 6.1 | 360 | 80 |

References for Spectral types:

1 = Keenan and McNeil (1989)

2 = Kirkpatrick, Henry and McCarthy (1991)

3 = BSC

Table 2

Band edges for continuum and features

| Feature Name | band edges $\lambda$ ($\mu$m) | | Transition | Lower State Energy (eV) |
|---|---|---|---|---|
| Continuum # 1 | 2.214 | 2.220 | – | – |
| Continuum # 2 | 2.233 | 2.237 | – | – |
| Continuum # 3 | 2.249 | 2.253 | – | – |
| Continuum # 4 | 2.255 | 2.258 | – | – |
| Continuum # 5 | 2.270 | 2.274 | – | – |
| Continuum # 6 | 2.287 | 2.290 | – | – |
| H I Brackett $\gamma$ | 2.160 | 2.170 | $4\ ^2F^o - 7^2G$ | 12.70 |
| NaI | 2.204 | 2.211 | $4s^2S_{1/2} - 4p^2P^o_{1/2}$ | 3.19 |
| NaI | 2.204 | 2.211 | $4s^2S_{1/2} - 4p^2P^o_{3/2}$ | 3.19 |
| CaI | 2.258 | 2.269 | $4d^3D_{3,2,1} - 4f^3F^o_4$ | 4.68 |
| CaI | 2.258 | 2.269 | $4d^3D_{3,2,1} - 4f^3F^o_3$ | 4.68 |
| CaI | 2.258 | 2.269 | $4d^3D_{3,2,1} - 4f^3F^o_2$ | 4.68 |
| MgI | 2.279 | 2.285 | $4d^3D_{3,2,1} - 6f^3F^o_{2,3,4}$ | 6.72 |
| $^{12}CO(2,0)$ | 2.289 | 2.302 | (2,0) bandhead | 0.32 |
| $H_2O$ red | 2.220 | 2.230 | – | – |
| $H_2O$ blue | 2.160 | 2.170 | – | – |

Table 3

Strengths of the target features

| Star | Spectral Type | Equivalent Width (Å) | | | | | | | | | | | | |
|---|---|---|---|---|---|---|---|---|---|---|---|---|---|---|
| | | Na I | | | Ca I | | | Mg I | | | $^{12}$CO(2,0) | | | F |
| BS 2353 | F3 V | -0.01 | ± | 0.33 | 0.93 | ± | 0.20 | 0.41 | ± | 0.20 | 0.45 | ± | 0.38 | |
| BS 147 | F6 V | 0.83 | ± | 0.21 | 0.34 | ± | 0.17 | 0.04 | ± | 0.07 | 0.57 | ± | 0.39 | |
| BS 2313 | F8 V | 0.83 | ± | 0.50 | 0.97 | ± | 0.34 | 0.30 | ± | 0.17 | 1.53 | ± | 0.13 | |
| BS 506 | F9 V | 1.01 | ± | 0.19 | 0.45 | ± | 0.19 | 0.62 | ± | 0.12 | 2.35 | ± | 0.30 | |
| BS 695 | G0 V | 1.28 | ± | 0.12 | 1.12 | ± | 0.35 | 0.21 | ± | 0.14 | 1.43 | ± | 0.08 | |
| HD 20619 | G1.5 V | -0.70 | ± | 0.29 | 0.41 | ± | 0.15 | 0.33 | ± | 0.17 | 0.76 | ± | 0.24 | |
| BS 2290 | G2 V | 0.72 | ± | 0.27 | 1.24 | ± | 0.35 | 0.20 | ± | 0.30 | 2.90 | ± | 0.31 | |
| BS 3538 | G2 V | 0.98 | ± | 0.26 | 1.46 | ± | 0.25 | 0.22 | ± | 0.31 | 3.32 | ± | 0.26 | |
| BS 88 | G2.5 V | 1.15 | ± | 0.21 | 1.10 | ± | 0.23 | 0.98 | ± | 0.19 | 1.88 | ± | 0.28 | |
| BS 4523 | G3 V | 0.88 | ± | 0.11 | 0.82 | ± | 0.10 | 1.27 | ± | 0.13 | 1.56 | ± | 0.16 | |
| BS 8631 | G4 V | 1.07 | ± | 0.09 | 1.33 | ± | 0.34 | 1.02 | ± | 0.27 | 3.85 | ± | 0.39 | |
| BS 996 | G5 V | 0.78 | ± | 0.10 | 1.73 | ± | 0.17 | 1.12 | ± | 0.05 | 1.52 | ± | 0.17 | |
| BS 5019 | G6.5 V | 1.27 | ± | 0.09 | 1.14 | ± | 0.10 | 1.90 | ± | 0.29 | 2.48 | ± | 0.31 | |
| BS 2692 | G9 V | 0.99 | ± | 0.14 | 1.09 | ± | 0.15 | 0.71 | ± | 0.25 | 5.29 | ± | 0.27 | |
| BS 166 | K0 V | 2.39 | ± | 0.28 | 2.07 | ± | 0.13 | 1.97 | ± | 0.11 | 4.30 | ± | 0.40 | |
| BS 493 | K1 V | 1.83 | ± | 0.14 | 1.93 | ± | 0.17 | 1.80 | ± | 0.09 | 3.78 | ± | 0.14 | |
| BS 1982 | K2 V | 2.10 | ± | 0.10 | 3.91 | ± | 0.19 | 2.25 | ± | 0.47 | 5.27 | ± | 0.92 | |
| BS 753 | K3- V | 2.58 | ± | 0.26 | 2.69 | ± | 0.27 | 1.68 | ± | 0.05 | 6.04 | ± | 0.26 | |
| BS 8721 | K4 V | 2.63 | ± | 0.07 | 3.47 | ± | 0.26 | 1.74 | ± | 0.06 | 4.93 | ± | 0.13 | |
| GL 204 | K5 V | 3.06 | ± | 0.06 | 3.56 | ± | 0.13 | 1.66 | ± | 0.05 | 7.04 | ± | 0.18 | |
| HD 120467 | K6 V | 3.06 | ± | 0.43 | 2.36 | ± | 0.64 | 0.95 | ± | 0.43 | 3.63 | ± | 0.12 | |
| GL 123 | M0 V | 4.26 | ± | 0.22 | 4.39 | ± | 0.18 | 1.90 | ± | 0.07 | 7.15 | ± | 0.17 | |
| GL 846 | M0.5 V | 4.58 | ± | 0.04 | 4.80 | ± | 0.14 | 1.63 | ± | 0.05 | 7.61 | ± | 0.12 | |
| GL 825 | M1 V | 3.96 | ± | 0.13 | 4.37 | ± | 0.41 | 0.92 | ± | 0.17 | 7.88 | ± | 0.29 | |
| HD 36395 | M1.5 V | 7.07 | ± | 0.09 | 6.82 | ± | 0.07 | 1.82 | ± | 0.18 | 9.05 | ± | 0.28 | |
| GL 382 | M2 V | 4.87 | ± | 0.12 | 5.87 | ± | 0.10 | 1.68 | ± | 0.74 | 6.44 | ± | 0.36 | |
| GL 250B | M2.5 V | 4.03 | ± | 0.17 | 5.10 | ± | 0.11 | 0.60 | ± | 0.20 | 7.34 | ± | 0.32 | |
| GL 352AB | M3 V | 5.22 | ± | 0.13 | 4.78 | ± | 0.07 | 1.50 | ± | 0.27 | 6.21 | ± | 0.87 | |
| GL 213 | M4 V | 4.18 | ± | 0.16 | 2.83 | ± | 0.06 | 1.02 | ± | 0.15 | 6.11 | ± | 0.57 | |
| GL 166C | M4.5 V | 3.81 | ± | 0.09 | 2.49 | ± | 0.33 | 0.46 | ± | 0.45 | 8.15 | ± | 0.35 | |
| GL 83.1 | M4.5 V | 4.64 | ± | 0.14 | 2.36 | ± | 0.08 | 0.45 | ± | 0.25 | 8.27 | ± | 0.21 | |
| GL 866AB | M5 V | 5.91 | ± | 0.19 | 1.47 | ± | 0.20 | 0.52 | ± | 0.32 | 11.76 | ± | 0.21 | |
| GL 406 | M6 V | 6.74 | ± | 0.17 | 1.88 | ± | 0.04 | 0.02 | ± | 0.23 | 13.28 | ± | 0.13 | |

Table 4
Fit Co-efficients

| | | | | | | | |
|---|---|---|---|---|---|---|---|
| Ca I (hot): | $T_{eff}$ | $=$ | 6559 | $\pm$ | 165 | | |
| | | $-$ | 748 | $\pm$ | 134 | $\times$ | EW |
| | | $+$ | 38.2 | $\pm$ | 20.6 | $\times$ | $EW^2$ |
| Ca I (cool): | $T_{eff}$ | $=$ | 2707 | $\pm$ | 96 | | |
| | | $-$ | 159 | $\pm$ | 22 | $\times$ | EW |
| Na I: | $T_{eff}$ | $=$ | 6397 | $\pm$ | 107 | | |
| | | $-$ | 630 | $\pm$ | 34 | $\times$ | EW |
| CO: | $T_{eff}$ | $=$ | 6651 | $\pm$ | 221 | | |
| | | $-$ | 482 | $\pm$ | 83 | $\times$ | EW |
| | | $+$ | 12.1 | $\pm$ | 6.6 | $\times$ | $EW^2$ |
| $H_2O$: | $T_{eff}$ | $=$ | 11937 | $\pm$ | 1686 | | |
| | | $-$ | 8034 | $\pm$ | 1574 | $\times$ | EW |

Table 5

Error in effective temperature determination

| Feature Name | $\pm\sigma_T$ (K) |
|---|---|
| NaI | 450 |
| CaI | 300 |
| $H_2O$ | 200 |
| $^{12}CO(2,0)$ | 450 |